\documentclass{thiselsart}

\usepackage{amssymb}
\input{epsf}
\begin{document}

\begin{frontmatter}



\title{A local hidden variable theory for the GHZ experiment}

\author{L\'aszl\'o E. Szab\'o}
\ead{leszabo@hps.elte.hu}
\address{Theoretical Physics Research Group of the Hungarian Academy of
 Sciences\\
Department of History and Philosophy of Science\\
E\"otv\"os University, Budapest, Hungary}

\author{Arthur Fine}
\ead{afine@u.washington.edu}
\address{Department of Philosophy\\University of Washington, Seattle, Washington
 98195-3550, USA}
\begin{abstract}
A recent analysis by de Barros and Suppes of experimentally realizable
GHZ correlations supports the conclusion that these correlations
cannot be explained by introducing local hidden variables. We show,
nevertheless, that their analysis does not exclude local hidden
variable models in which the inefficiency in the experiment is an
effect not only of random errors in the detector
equipment, but is also the manifestation of a pre-set, hidden
property of the particles (``prism models''). Indeed, we present an
explicit prism model for the GHZ scenario; that is, a local hidden
variable model entirely compatible with recent GHZ experiments.

\end{abstract}

\end{frontmatter}


\section{Introduction}

De Barros and Suppes \cite{BS}  give a general analysis of realistic
experiments, where experimental error reduces the perfect
correlations of the ideal GHZ case. Their analysis makes use of
inequalities which are said to be ``both necessary and sufficient for
the existence of a local hidden variable'' for the experimentally
realizable GHZ correlations. In applying their analysis to the
Innsbruck experiment \cite{Z}, however, they only count events in
which
all the detectors fire. While necessary for the analysis of that
experiment, they recognize that this selective procedure weakens the
argument for the non-existence of local hidden variables. Here we
show that they are right and that their analysis does not
rule out a whole class of local hidden variable models in which the
detection inefficiency is not (only) the effect of the random errors in
the detector equipment, but it is a more fundamental phenomenon, the
 manifestation
of a predetermined hidden property of the particles. This conception of 
local hidden variables was suggested in Fine's \emph{prism model} \cite{Fa}
and, arguably, goes back to Einstein (See \cite{Game} Chapter 4).
 
Prism models work well in case of the EPR--Bell experiments. The
original model applied to the \( 2\times 2 \)
spin-correlation experiments and was in complete accordance with
the known experimental results. There appeared, however, a
theoretical demand to embed the \( 2\times 2 \) prism models into a
large \( n\times n \) prism model reproducing all potential \(
2\times 2 \) sub-experiments. This demand was motivated by the idea
that the real physical process does not know which directions are
chosen in an experiment. On the other hand, it seemed that in the
known prism models of the \( n\times n \) spin-correlation experiment
the efficiencies tended to zero, if \( n\rightarrow \infty  \), which
contradicts what we expect of actual experiments. This contradiction was
recently resolved in \cite{Larsson2} and \cite{Sz}, which show that there is a
 wide
class of physically plausible \( \infty\times \infty \) prism models
with high efficiency ($\leq 0.82$).

In the first part of this paper we explain the principle difference
between the prism models and the local hidden variable models to
which de Barros and Suppes' analysis applies. In the second part, we
present an explicit prism model for the GHZ scenario, a local hidden
variable model that is entirely compatible with recent GHZ
experiments.

\section{The GHZ experiment}

\begin{figure}[t]
	\begin{center}\leavevmode
	\epsfxsize=6cm
	\epsfbox{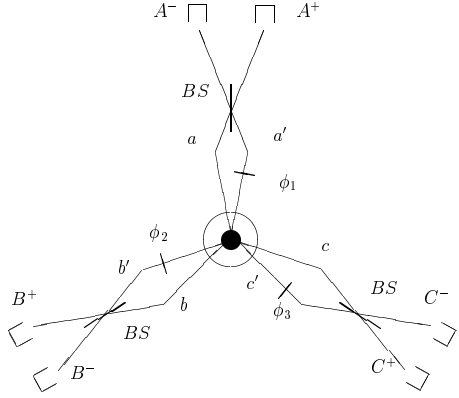}
	\end{center}
	\caption{A three-particle beam-entanglement interferometer}
\end{figure}

Greenberger, Horne, Shimony and Zeilinger \cite{GHSZ} developed a proof
of
the Bell theorem without using inequalities. For the GHZ example
consider three entangled photons flying apart along three different
straight lines in the horizontal plane (Fig.~1). Assume that the
(polarization part of the) quantum state of the three-photon system is

\begin{equation}
\Psi =\frac{1}{\sqrt{2}}\left( \left| H\right\rangle _{1}\otimes
\left| H\right\rangle _{2}\otimes \left| V\right\rangle _{3}+\left|
V\right\rangle _{1}\otimes \left| V\right\rangle _{2}\otimes \left|
H\right\rangle _{3}\right)
\end{equation}

One can transform the polarization degree of freedom into the momentum
degree of freedom by means of polarizing beam splitters (see \cite{ZHWZ}). 
So the quantum state of the system can be written also in the
following form:
  \[
\Psi =\frac{1}{\sqrt{2}}\left( \left| a\right\rangle _{1}\otimes
\left| b\right\rangle _{2}\otimes \left| c\right\rangle _{3}+\left|
a'\right\rangle _{1}\otimes \left| b'\right\rangle _{2}\otimes \left|
c'\right\rangle _{3}\right) \]
where \( \left| a\right\rangle _{1} \) denotes the particle 1 in beam
$a$, etc. A straightforward interferometric calculation (\cite{GHSZ}, p. 1141)
 shows that the probabilities of detections are
\begin{equation}
\label{valoszinusegek}
\begin{array}{rcl}
p^{\Psi }_{A^{+}B^{+}C^{+}}\left( \phi _{1},\phi _{2},\phi _{3}\right)  & = &
 \frac{1}{8}\left( 1+\sin \left( \phi _{1}+\phi _{2}+\phi _{3}\right) \right) \\
p^{\Psi }_{A^{-}B^{+}C^{+}}\left( \phi _{1},\phi _{2},\phi _{3}\right)  & = &
 \frac{1}{8}\left( 1-\sin \left( \phi _{1}+\phi _{2}+\phi _{3}\right) \right) \\
 & \textrm{etc}. & 
\end{array}
\end{equation}
(If the number of minuses on the detector labels is even, there is a plus sign;
 if odd, there is a minus sign.)
Introduce the following result functions
\[
A\left( \phi _{1}\right) =\left\{ \begin{array}{ll}
1 & {\textrm{if the detector }} A^{+}\textrm{ fires}\\
-1 & {\textrm{if the detector }}A^{-}\textrm{ fires}
\end{array}\right. \]
\noindent \( B\left( \phi _{2}\right)  \) and \( C\left( \phi _{3}\right)
\)
have the same meaning for particles \( 2 \) and \( 3 \). One can also show
that  in state \( \Psi  \) the expectation value of the product of the three
 outcomes is
\[
E\left( A\left( \phi _{1}\right) B\left( \phi _{2}\right) C\left( \phi
 _{3}\right) \right) =\sin \left( \phi _{1}+\phi _{2}+\phi _{3}\right) \]
Consider the following choices of angles:
\begin{equation}
\begin{array}{rcl}
\Omega _{1} & = & A\left( \pi /2\right) B\left( 0\right) C\left(
0\right) \\
\Omega _{2} & = & A\left( 0\right) B\left( \pi /2\right) C\left(
0\right) \\
\Omega _{3} & = & A\left( 0\right) B\left( 0\right) C\left( \pi
/2\right)  \\
\Omega _{4} & = & A\left( \pi /2\right) B\left( \pi /2\right) C\left(
\pi /2\right) 
\end{array}
\end{equation} 
In this case we obtain perfect correlations:
\begin{eqnarray}
E(\Omega _{1})=E(\Omega _{2})=E(\Omega _{3}) & = & 1\label{expectation1}
\\
E(\Omega _{4}) & = & -1\label{expectation2}
\end{eqnarray}

So far this is standard quantum mechanics. One can make a
Kochen--Specker/EPR-type
argument, however, if one assumes that in \( \Psi  \) predetermined
values, revealed by measurement, are assigned to the six observables
\begin{equation}
A\left( \pi /2\right) ,A\left( 0\right) ,B\left( \pi /2\right)
,B\left( 0\right) ,C\left( \pi /2\right) ,C\left( 0\right) \label{combination}
\end{equation}

By virtue of (\ref{expectation1}) and (\ref{expectation2}) these values have to
 satisfy the following constraints:
\begin{equation}
\label{constraints}
\begin{array}{rcccl}
\Omega _{1} & = & A\left( \pi /2\right) B\left( 0\right) C\left(
0\right)  & = & 1\\
\Omega _{2} & = & A\left( 0\right) B\left( \pi /2\right) C\left(
0\right)  & = & 1\\
\Omega _{3} & = & A\left( 0\right) B\left( 0\right) C\left( \pi
/2\right)  & = & 1\\
\Omega _{4} & = & A\left( \pi /2\right) B\left( \pi /2\right) C\left(
\pi /2\right)  & = & -1
\end{array}
\end{equation} 

Then a contradiction is immediate if we take the product of equations
(\ref{constraints}). Each value appears twice so, whatever the assigned values
 are,
the left hand side is a positive number, whereas the right side is \(
-1 \).

\section{De Barros and Suppes' inequalities}

De Barros and Suppes approach the above
contradiction in the
following way. Without loss of generality, the space of hidden variable
can be identified with
  ${\mathcal{O}}=\{+,-\}^6$, the set of the \( 2^{6}=64
\) different 6-tuples of possible combinations of the values (6).
Then the GHZ contradiction amounts to the
assertion that no probability measure over \( \mathcal{O}
\) reproduces the expectation values (\ref{expectation1}) and
(\ref{expectation2}).
	De Barros and Suppes demonstrate this by concentrating on the
product observables ($\Omega_1, \dots, \Omega_4$) for which they derive a
system
of
inequalities that play the same role for GHZ that the general form of
the Bell inequalities do for EPR-Bohm type experiments \cite{Fb};
namely, they provide necessary and sufficient conditions for a
certain class of local hidden variable models. The first of their
inequalities is just
\[ -2  \leq   E(\Omega _{1})+E(\Omega _{2})+E(\Omega
_{3})-E(\Omega_{4})  \leq  2 \]
and clearly this is violated by (\ref{expectation1}) and
(\ref{expectation2}). Moreover if, due to inefficiencies in the
detectors or to  dark photon detection, the observed correlations
were reduced by some factor $\varepsilon$; that is
   \begin{eqnarray}
E(\Omega _{1})=E(\Omega _{2})=E(\Omega _{3}) & = & 1-\varepsilon
\label{expectation3} \\
E(\Omega _{4}) & = & -1+\varepsilon \label{expectation4}
\end{eqnarray}
then, it follows immediately from this inequality that, ``the observed
correlations are only compatible with a local hidden variable theory''
if \( \varepsilon >\frac{1}{2} \). De Barros and Suppes made a detailed analysis
 of the detection errors  and the dark photon detections in the realistic
 detector equipments, and concluded that these phenomena do not yield such a
 large $\varepsilon$.

	As in the case of the Bell inequalities, however, the de
Barros and Suppes derivation starts with the assumption
that the variables in (6) are two valued (either $+1$ or $-1$). Since we
 consider the detection/emission ratio as of more fundamental origin, in the
 prism models developed in the next sections, the variables can take on a third
 value,``$D$'', corresponding to an inherent ``no show'' or
defectiveness. In the Bell-EPR case we know that the existence of
local hidden variables of this more general type are governed
by a different system of inequalities. For the inversion symmetric
2x2 case inequalities providing necessary and sufficient conditions
for prism models were derived in \cite{GM}. We do not
have a comparable system characterizing prism models for GHZ type
experiments but we will show that GHZ experiments can be modeled by
just such local hidden variable theories. Indeed we will give an
explicit prism model for a GHZ experiment (with perfect detectors and with zero
 dark-photon detection probability). We will also show that our model is
completely compatible with the results measured in the Innsbruck
experiment.

\section{A toy prism model of the GHZ experiment}
\label{toy}
The prism model of the GHZ experiment is a local, deterministic
hidden variable theory, in which the hidden variables predetermine
not only the outcomes of the corresponding measurements, but also
predetermine whether or not an emitted particle arrives to the
detector and becomes detected. Consequently, the space $\Lambda$ of
hidden variables ought to be a subset of $\{+,-,D\}^6$.  Each
element of $ \Lambda$ is a 6-tuple that corresponds to combinations
like \[ \left( A(\pi /2),A(0),B(\pi /2),B(0),C(\pi /2) ,C( 0) \right)
=(+-D-++) \] which, for example, stands for the case when particle 1
is predetermined to produce the outcome \( +1 \) if \( \phi _{1}=\pi
/2 \), \( -1 \) if angle \( \phi _{1}=0 \) in the measurement,
particle 2 is \( \pi /2 \)-defective, i.e., it gives no outcome if \(
\phi _{2}=\pi /2 \), but produces an outcome \( -1 \) if \( \phi
_{2}=0 \), particle 3 produces outcome \( +1 \) for both cases. The
essential feature of this conception of hidden variables is that
the ``values'' \( A_{\lambda }(\pi /2),A_{\lambda }(0),B_{\lambda
}(\pi /2),\dots  \) are ``prismed'' in the sense that, formally, a
new ``value'' is introduced, ``\( D \)'', corresponding to the case
when the particle is predetermined not to produce an outcome. 

Each GHZ event will be represented as a subset $U$ of $\Lambda$. For instance \[
U_{\{A(\pi/2)=+\}\&\{B(0)=-\}\&\{C(0)=-\}} \] stands for the triple detection 
 $A^{+}B^{-}C^{-}$ with angles $(\pi/2, 0,0)$.

We have seen that, if
determinate values are assigned to all the observables, quantum
mechanics yields contradictory correlations (\ref{constraints}) among the
 measurement
outcomes at the three stations. Although these four correlations are enough for
 the stated GHZ contradiction, our hidden variable model must be consistent with
 quantum mechanics in a  wider sense: the probability measure $p$ we are going
 to define on $\Lambda \subset \{+,-,D\}^6$ must satisfy further constraints,
 following from (\ref{valoszinusegek}):
\begin{equation}
\label{QM}
\begin{array}{lcr}
  p\left( U_{\{A(x)=i\}\&\{B(y)=j\}\&\{C(z)=k\}}|U^{triple}_{(x,y,z)}\right)
 &\\=\frac{1}{8}\left[ 1+ijk\sin (x+y+z)\right]  & \\
 \ \ \ \ \ \ \ \ \ \ \ \ \ \ \ \ \ \ \ \ \ \ \ \ \  \textrm{for all
 }\begin{array}{rcl}
x,y,z & = & \frac{\pi }{2},0\\
i,j,k & = & \pm 1
\end{array} &
\end{array}
\end{equation}
where \( U^{triple}_{(x,y,z)}=U_{\{A(x)\neq D\}\&\{B(y)\neq D\}\&\{C(z)\neq D\}}
 \), the event of triple detection. 

Constraints (\ref{constraints}) correspond to the fact that some of these
 probabilities are zero, which rule  out a large number of 6-tuples.  One can
 show (and easily verify
by computer) that from the $3^6=729$ elements of $\{+,-,D\}^6$ there
remains 409 which satisfy (\ref{constraints}). For example:

\begin{itemize}
\item $(-++-DD)$ is allowed, because, in this case,
whatever the chosen experimental setup, there is no detection at
station 3, consequently there is no triple coincidence detection.
\item $(D--DD+)$ is allowed because for any measurement setup either
the outcome triad satisfies the constraints or there is no triple
coincidence at all. 
\item \( (---D+-)  \) is not allowed, because if the chosen angles were
$\left(\pi/2,\pi/2,\pi/2 \right)$ then the results would be $A(\pi/2) =
-$, $B(\pi /2)=-$ and $C(\pi/2) = +$, which
would contradict the constraint $\Omega _{4}=-1$.
\end{itemize}

There is a prism model on the hidden variable space consisting of
these left 409 elements. However, in order to achieve better
detection/emission efficiencies, and also to simplify the model, we
will refine $\Lambda$ further. The 409 combinations form four
disjoint subsets: 217 of them correspond to the situation where there
is no triple detection at all, regardless of the angles chosen at the
three stations; 48 combinations produce a triple detection
coincidence at only one, and 96 at two triads of angles (these 48 and 96 form a
 prism model for GHZ all by themselves) and the remaining 48 combinations
 produce a
triple coincidence with four different triads of experimental setups.
Clearly we achieve the best efficiency if we take for $\Lambda$ the
fourth subset, listed in Table~I, and simply omit all the others.%
\footnote{See \cite{Larsson0}  and \cite{Larsson1} for a different presentation
 of the  48-model, and
also for important calculations concerning error bounds for GHZ type
experiments.}
\begin{table}[t]
\[ \begin{array}{rcl}
\lambda_{ 1} $=$ ( - - - - D  +   )\\
\lambda_{ 2} $=$ ( - - - D  - +   )\\
\lambda_{ 3} $=$ ( - - - + - D    )\\
\lambda_{ 4} $=$ ( - - D  - + +   )\\
\lambda_{ 5} $=$ ( - - D  + - -   )\\
\lambda_{ 6} $=$ ( - - + - + D    )\\
\lambda_{ 7} $=$ ( - - + D  + -   )\\
\lambda_{ 8} $=$ ( - - + + D  -   )\\
\lambda_{ 9} $=$ ( - D  - - - +   )\\
\lambda_{ 10} $=$ ( - D  - + - -   )\\
\lambda_{ 11} $=$ ( - D  + - + +   )\\
\lambda_{ 12} $=$ ( - D  + + + -   )\\
\lambda_{ 13} $=$ ( - + - - - D    )\\
\lambda_{ 14} $=$ ( - + - D  - -   )\\
\lambda_{ 15} $=$ ( - + - + D  -   )\\
\lambda_{ 16} $=$ ( - + D  - - +   )\\
\lambda_{ 17} $=$ ( - + D  + + -   )\\
\lambda_{ 18} $=$ ( - + + - D  +   )\\
\lambda_{ 19} $=$ ( - + + D  + +   )\\
\lambda_{ 20} $=$ ( - + + + + D    )\\
\lambda_{ 21} $=$ ( D  - - - + +   )\\
\lambda_{ 22} $=$ ( D  - - + - +   )\\
\lambda_{ 23} $=$ ( D  - + - + -   )\\
\lambda_{ 24} $=$ ( D  - + + - -   )\\
\end{array}\ \ \ 
\begin{array}{rcl}
\lambda_{ 25} $=$ ( D  + - - - -   )\\
\lambda_{ 26} $=$ ( D  + - + + -   )\\
\lambda_{ 27} $=$ ( D  + + - - +   )\\
\lambda_{ 28} $=$ ( D  + + + + +   )\\
\lambda_{ 29} $=$ ( + - - - + D    )\\
\lambda_{ 30} $=$ ( + - - D  + +   )\\
\lambda_{ 31} $=$ ( + - - + D  +   )\\
\lambda_{ 32} $=$ ( + - D  - + -   )\\
\lambda_{ 33} $=$ ( + - D  + - +   )\\
\lambda_{ 34} $=$ ( + - + - D  -   )\\
\lambda_{ 35} $=$ ( + - + D  - -   )\\
\lambda_{ 36} $=$ ( + - + + - D    )\\
\lambda_{ 37} $=$ ( + D  - - + -   )\\
\lambda_{ 38} $=$ ( + D  - + + +   )\\
\lambda_{ 39} $=$ ( + D  + - - -   )\\
\lambda_{ 40} $=$ ( + D  + + - +   )\\
\lambda_{ 41} $=$ ( + + - - D  -   )\\
\lambda_{ 42} $=$ ( + + - D  + -   )\\
\lambda_{ 43} $=$ ( + + - + + D    )\\
\lambda_{ 44} $=$ ( + + D  - - -   )\\
\lambda_{ 45} $=$ ( + + D  + + +   )\\
\lambda_{ 46} $=$ ( + + + - - D    )\\
\lambda_{ 47} $=$ ( + + + D  - +   )\\
\lambda_{ 48} $=$ ( + + + + D  +   )
\end{array} \]
\caption{ }
\end{table}

For instance the event ``\(
B(0)=+ \)'' corresponds to
\begin{eqnarray*}
U_{\{B(0)=+\}}=\left( \lambda_{3},\lambda_{5}, \lambda_{8},
\lambda_{10}, \lambda_{12},
\lambda_{15}, \lambda_{17}, \lambda_{20}, \lambda_{22}, \lambda_{24},\right.\\
\left.\lambda_{26},
\lambda_{28}, \lambda_{31}, \lambda_{36}, \lambda_{38}, \lambda_{40},
 \lambda_{43},\lambda_{45},\lambda_{48}\right)
\end{eqnarray*}
\noindent Similarly, the event, for example, that ``\( A(\pi /2)=+ \)
and \( B(0)=+ \)'' is represented by the following subset:
\begin{eqnarray*}
U_{\{A(\pi/2)=+\}\&\{B(0)=+\}}= \left(\lambda_
{31}, \lambda_{33}, \lambda_{36}, \lambda_{38}, \lambda_{40}, \lambda_{43},
 \lambda_{45}, \lambda_{48}\right)
\end{eqnarray*}
while, for instance, $
U_{\{A(\pi/2)=+\}\&\{B(0)=+\}\&\{C(0)=-\}}=\emptyset$.

Notice that each subset $U_{\{A(x)\neq D\}\&\{B(y)\neq D\}\&\{C(z)\neq
D\}}$ --  where $x,y,z=\pi/2$ or $0$ -- consists of exactly 24 
elements of \( \Lambda  \). These subsets correspond to the triple
measurement events that enter into GHZ.

The probability measure on $\Lambda$ must be defined in such a way that the rest
 of conditions  (\ref{QM}) (right hand side is not zero) be satisfied. One can
 verify by computer that the uniform
distribution on \( \Lambda  \) is a suitable one (each element has
probability $\frac{1}{48}$) and the probability model $\left(\Lambda,
p\right)$ thus obtained has maximal triple detection efficiency.
Indeed, the triple efficiencies are:
\begin{eqnarray*}
p(\mathrm{triple\ coincidence})&=&p\left( U_{\{A(x)\neq
D\}\&\{B(y)\neq D\}\&\{C(z)\neq D\}}\right)=\frac{24}{48}=0.5
\end{eqnarray*}
The only way to increase the efficiency would be to modify the
probability distribution over $\Lambda$. Assuming, however, that for
such a non-uniform distribution the triple coincidence efficiency is
still independent of the chosen experimental setups, we have
\begin{eqnarray*}
p(\mathrm{triple\ coincidence})&=&\sum_{i=1}^{48} p({\mathrm{triple\
coincidence}}|\lambda_i)p(\lambda_i)\\&=&\sum_{i=1}^{48} \frac{4}{2^3}
p(\lambda_i)= \frac{1}{2}
\end{eqnarray*}
independently of the actual probability distribution $p(\lambda_i)$.

The key idea of a prism model now is to retrieve the quantum
probabilities $q(.)$ as the \( \Lambda  \) space probabilities
conditional on the measurement outcomes being nondefective. Due to conditions
 (\ref{QM}) this feature of the model is automatically provided. Assume,
for example, that the chosen angles are \( \left\{ \pi /2,0,0\right\}  \), then
\begin{eqnarray*}
& &q\left( \{A(\pi/2)=-\}\right)\\
&=&p\left(U_{\{A(\pi/2)=-\}}|U_{\{A(\pi/2)\neq D\}\&\{B(0)\neq
D\}\&\{C(0)\neq D\}}\right)\\
&=&\frac{p\left(U_{\{A(\pi/2)=-\}}\cap U_{\{A(\pi/2)\neq D\}\&\{B(0)\neq
D\}\&\{C(0)\neq D\}}\right)}{p\left( U_{\{A(\pi/2)\neq D\}\&\{B(0)\neq
D\}\&\{C(0)\neq D\}}\right)}\\
&=&\frac{p\left(\left\{\lambda_{1}, \lambda_{4}, \lambda_{5},
\lambda_{8}, \lambda_{9},
\lambda_{10}, \lambda_{11}, \lambda_{12}, \lambda_{15}, \lambda_{16},
 \lambda_{17},
 \lambda_{18}\right\}\right)}{\frac{24}{48}}\\&=&\frac{\frac{12}{48}}{\frac{
24}{48}}=\frac{1}{2}
\end{eqnarray*}
Similarly, all the other observed single detection probabilities  are
 $\frac{1}{2}$.

To illustrate that constraints (\ref{QM}) are satisfied, consider, for
example,
\begin{eqnarray*}
& &q\left(\{A(\pi/2)=+\}\&\{B(\pi/2)=+\}\&\{C(0)=-\}\right)\\
&=&p\left(U_{\{A(\pi/2)=+\}}\cap U_{\{B(\pi/2)=+\}}\cap
U_{\{C(0)=-\}}| U^{triple}_{(\frac{\pi}{2},\frac{\pi}{2},0)} \right)\\
&=&\frac{p\left(U_{\{A(\pi/2)=+\}}\cap U_{\{B(\pi/2)=+\}}\cap
U_{\{C(0)=-\}}\cap
 U^{triple}_{(\frac{\pi}{2},\frac{\pi}{2},0)}\right)}{p\left(U^{triple}_{(\frac{
\pi}{2},\frac{\pi}{2},0)}\right)}\\
&=&\frac{p\left(\left\{\lambda_{34},\lambda_{35},\lambda_{39}\right\}\right)}{\frac{24}{48}}=\frac{\frac{3}{48}}{\frac{24}{48}}=\frac{1}{8}\\&=&\frac{1}{8}\left
[ 1-\sin\left(\frac{\pi}{2}+\frac{\pi}{2}+0\right)\right]
\end{eqnarray*}
Finally, due to the selections involved in building the hidden
variable space $\Lambda$  the model correctly reproduces the GHZ
correlations (\ref{constraints}), whenever a
triple detection coincidence occurs: For
example the {\em observed} 
expectation value of \(
\Omega _{1}
\)
is
\begin{eqnarray*}
E(\Omega _{1}) &=& \sum_{\lambda \in
 U^{triple}_{(\frac{\pi}{2},0,0)}}\Omega_1(\lambda)p(\lambda|
U^{triple}_{(\frac{\pi}{2},0,0)})=
\sum_{\lambda \in
 U^{triple}_{(\frac{\pi}{2},0,0)}}\frac{p(\lambda)}{p\left(U^{triple}_{(\frac{%
 \pi}{2},0,0)}\right)}=1
\end{eqnarray*}
\noindent and, similarly,
\begin{eqnarray*}
E(\Omega _{2})=E(\Omega _{3}) & = & 1\\
E(\Omega _{4}) & = & -1
\end{eqnarray*}

According to the key idea of a prism model, the above expectation
values are calculated on sub-ensembles of the emitted particle triads
that produce triple detection coincidences. In this respect the prism
model mirrors actual GHZ experiments.

\section{A complete infinite prism model for the GHZ experiment}

In the derivation of the GHZ contradiction we consider only \( 2\times 2\times 2
 \)
different experimental setups: At each station one considers two possible phase
shift angles: \( \phi _{1}=\frac{\pi }{2},0 \); \( \phi _{2}=\frac{\pi }{2},0
 \);
\( \phi _{3}=\frac{\pi }{2},0 \). In reality, however, the three angles can
be chosen arbitrarily, and, according to quantum mechanics, the resulted triple
detection probabilities are given in (\ref{valoszinusegek}). Although the particular
\( 2\times 2\times 2 \) scenario suffice to derive a negative statement, the
alleged contradiction between the existence of a local hidden variable theory
and the observed triple detection probabilities, it is not sufficient for a
positive statement about the existence of such a local hidden variable theory.
The reason is that if nature works according to such a local hidden variable
model, then the model must reproduce all observed triple detection probabilities
for all possible combinations of angles \( \phi _{1},\phi _{2},\phi _{3} \),
since the underlying physical process does not know about the  angles
chosen by the laboratory assistants at the three stations. That is why we call a toy model the one we constructed in the previous section, and the existence of
a complete infinite prism model covering the whole GHZ scenario is still an
open question. 

Now we are going to construct a complete infinite prism model for the GHZ experiment,
which covers the continuum case, when the three phase shift angles can take
arbitrary values. For the sake of later convenience introduce the following
new parametrization of the phase shift angles:\[
\begin{array}{ccc}
\alpha  & = & \phi _{1}-\frac{\pi }{6}\\
\beta  & = & \phi _{2}-\frac{\pi }{6}\\
\gamma  & = & \phi _{3}-\frac{\pi }{6}
\end{array}\]

Let us sum up what must be represented in the model: 
\begin{itemize}
\item [I.] A continuum set of events corresponding to the detection events,\[
A^{+}_{\alpha },A^{-}_{\alpha },B^{+}_{\beta },B^{-}_{\beta },C^{+}_{\gamma },C^{-}_{\gamma }\, \, \, \, \, \alpha ,\beta ,\gamma \in [0
,2\pi ]\]
together with the triple conjunctions\[
\begin{array}{c}
A^{+}_{\alpha }\wedge B^{+}_{\beta }\wedge C_{\gamma }^{+}\\
A^{+}_{\alpha }\wedge B^{+}_{\beta }\wedge C_{\gamma }^{-}\\
\vdots \\
A^{-}_{\alpha }\wedge B^{-}_{\beta }\wedge C_{\gamma }^{-}
\end{array}\, \, \, \, \alpha ,\beta ,\gamma \in [0,2\pi ]\]
And the {}``non-defectiveness{}'' events \begin{equation}
\label{eq_detections}
\begin{array}{rcl}
A_{\alpha } & = & A^{+}_{\alpha }\vee A^{-}_{\alpha }\\
B_{\beta } & = & B^{+}_{\beta }\vee B^{-}_{\beta }\\
C_{\gamma } & = & C^{+}_{\gamma }\vee C^{-}_{\gamma }
\end{array}\, \, \, \, \alpha ,\beta ,\gamma \in [0,2\pi ]
\end{equation}
together with the algebraic relations following from (\ref{eq_detections}).
\item [II.] The {}``quantum{}'' probabilities of the above events, 
\begin{eqnarray}
\label{eq_prob1}
\frac{p\left( A^{+}_{\alpha }\right) }{p\left( A_{\alpha }\right) }=\frac{p\left( A^{-}_{\alpha }\right) }{p\left( A_{\alpha }\right) }=\frac{p\left( B^{+}_{\beta }\right) }{p\left( B_{\beta }\right) }=\frac{p\left( B^{-}_{\beta }\right) }{p\left( B_{\beta }\right) }\nonumber\\
=\frac{p\left( C_{\gamma }^{+}\right) }{p\left( C_{\gamma }\right) }=\frac{p\left( C_{\gamma }^{-}\right) }{p\left( C_{\gamma }\right) }=\frac{1}{2}
\end{eqnarray}
\begin{eqnarray}
\label{eq_prob2}
\frac{p\left( A^{+}_{\alpha }\wedge B^{+}_{\beta }\wedge C_{\gamma }^{+}\right) }{p\left( A_{\alpha }\wedge B_{\beta }\wedge C_{\gamma }\right) }=\frac{p\left( A^{-}_{\alpha }\wedge B^{-}_{\beta }\wedge C_{\gamma }^{+}\right) }{p\left( A_{\alpha }\wedge B_{\beta }\wedge C_{\gamma }\right) }\nonumber\\
\frac{p\left( A^{-}_{\alpha }\wedge B^{+}_{\beta }\wedge C_{\gamma }^{+}\right) }{p\left( A_{\alpha }\wedge B_{\beta }\wedge C_{\gamma }\right) }=
\frac{p\left( A^{+}_{\alpha }\wedge B^{-}_{\beta }\wedge C_{\gamma }^{-}\right) }{p\left( A_{\alpha }\wedge B_{\beta }\wedge C_{\gamma }\right) }\nonumber\\
=\frac{1}{8}\left( 1-\cos \left( \alpha +\beta +\gamma \right) \right)
\end{eqnarray}
\begin{eqnarray}
\label{eq_prob3}
 \frac{p\left( A^{-}_{\alpha }\wedge B^{+}_{\beta }\wedge C_{\gamma }^{+}\right) }{p\left( A_{\alpha }\wedge B_{\beta }\wedge C_{\gamma }\right) }=\frac{p\left( A^{+}_{\alpha }\wedge B^{-}_{\beta }\wedge C_{\gamma }^{+}\right) }{p\left( A_{\alpha }\wedge B_{\beta }\wedge C_{\gamma }\right) }\nonumber\\
\frac{p\left( A^{+}_{\alpha }\wedge B^{+}_{\beta }\wedge C_{\gamma }^{-}\right) }{p\left( A_{\alpha }\wedge B_{\beta }\wedge C_{\gamma }\right) }=
\frac{p\left( A^{-}_{\alpha }\wedge B^{-}_{\beta }\wedge C_{\gamma }^{-}\right) }{p\left( A_{\alpha }\wedge B_{\beta }\wedge C_{\gamma }\right) }\nonumber\\
=\frac{1}{8}\left( 1+\cos \left( \alpha +\beta +\gamma \right) \right)
\end{eqnarray}
 in accordance with (\ref{valoszinusegek}). Probabilities \( p\left( A_{\alpha }\right)  \),
\( p\left( B_{\beta }\right)  \), \( p\left( C_{\gamma }\right)  \) and \( p\left( A_{\alpha }\wedge B_{\beta }\wedge C_{\gamma }\right)  \)
are the {}``single and triple detection efficiencies{}''.
\item [III.] The obvious symmetry of the whole experimental setup: None of the three
stations is privileged. That is, \begin{equation}
\label{eq_S1}
p\left( A_{\alpha }\right) =p\left( B_{\beta }\right) =p\left( C_{\gamma }\right) =\omega =\textrm{constant}
\end{equation}
\begin{eqnarray}
\label{eq_S2}
p\left( A_{\alpha }\wedge B_{\beta }\wedge C_{\gamma }\right) =p\left( A_{\beta }\wedge B_{\alpha }\wedge C_{\gamma }\right)\nonumber \\ =p\left( A_{\gamma }\wedge B_{\beta }\wedge C_{\alpha }\right) =\ldots 
\end{eqnarray}
\end{itemize}
\begin{figure}[t]
  	\begin{center}\leavevmode
	\epsfxsize=8cm
	\epsfbox{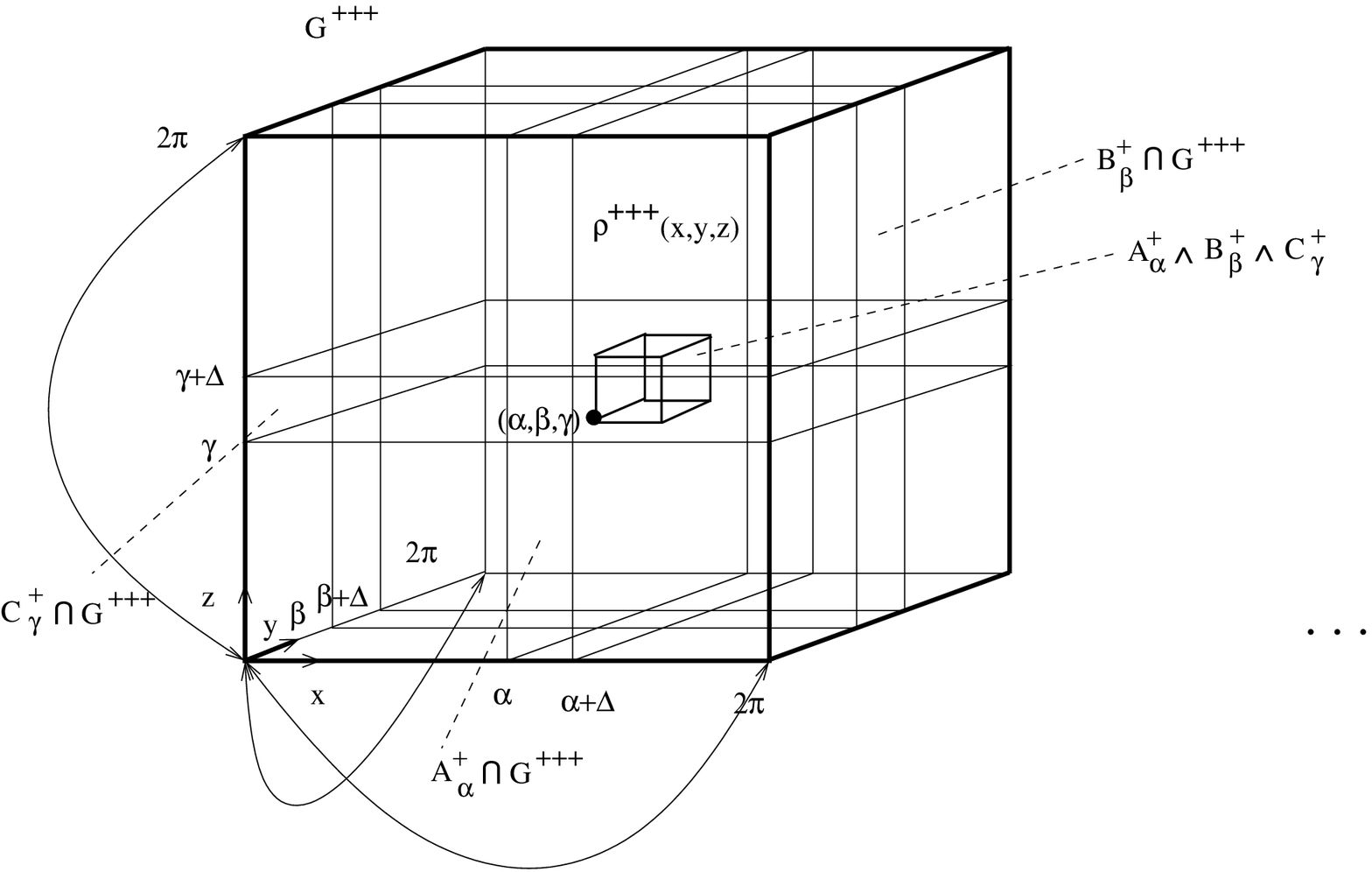}
  	\end{center}
	\caption{The hidden variable space is the union of eight regions \protect\( G^{+++}\protect \),
\protect\( G^{++-}\protect \),...\protect\( G^{---}\protect \). The first
one is shown in the figure}
\label{fig_space}
\end{figure}

Now, the local hidden variable model we are going to construct is based on the hidden variable space consisting of eight regions \( G^{+++} \), \( G^{-++} \),
\( G^{+-+} \) ,... \( G^{---} \). Each one is a space $S^1\times S^1\times S^1 $
, represented by a cube  of size \( (2\pi)\times (2\pi )\times (2\pi ) \), in which the points of coordinate $0$ and $2\pi$ are identified.(The first such region \( G^{+++} \) is shown in Fig.~\ref{fig_space}.) The normalized probability measure is given by the eight non-negative
density functions \( \rho ^{+++},\ldots \rho ^{---} \), such that\begin{eqnarray*}
\int _{0}^{2\pi }\int _{0}^{2\pi  }\int _{0}^{2\pi }\rho ^{+++}(x,y,z)dxdydz &  & \\
+\int _{0}^{2\pi  }\int _{0}^{2\pi  }\int _{0}^{2\pi  }\rho ^{-++}(x,y,z)dxdydz &  & \\
\ldots +\int _{0}^{2\pi  }\int _{0}^{2\pi }\int _{0}^{2\pi  }\rho ^{---}(x,y,z)dxdydz & = & 1
\end{eqnarray*}
The events are represented in the following way: \begin{eqnarray*}
A_{\alpha }^{+} & = & \left\{ (x,y,z)\in G^{+++}\left| \alpha \leq x\leq \alpha +\Delta \right. \right\} \\
 & \cup  & \left\{ (x,y,z)\in G^{+-+}\left| \alpha \leq x\leq \alpha +\Delta \right. \right\} \\
 & \cup  & \left\{ (x,y,z)\in G^{+--}\left| \alpha \leq x\leq \alpha +\Delta \right. \right\} \\
 & \cup  & \left\{ (x,y,z)\in G^{+--}\left| \alpha \leq x\leq \alpha +\Delta \right. \right\} \\
A_{\alpha }^{-} & = & \left\{ (x,y,z)\in G^{-++}\left| \alpha \leq x\leq \alpha +\Delta \right. \right\} \\
 & \cup  & \left\{ (x,y,z)\in G^{-+-}\left| \alpha \leq x\leq \alpha +\Delta \right. \right\} \\
 & \cup  & \left\{ (x,y,z)\in G^{--+}\left| \alpha \leq x\leq \alpha +\Delta \right. \right\} \\
 & \cup  & \left\{ (x,y,z)\in G^{---}\left| \alpha \leq x\leq \alpha +\Delta \right. \right\} \\
\vdots  &  & \\
C_{\gamma }^{-} & = & \left\{ (x,y,z)\in G^{++-}\left| \gamma \leq z\leq \gamma +\Delta \right. \right\} \\
 & \cup  & \left\{ (x,y,z)\in G^{-+-}\left| \gamma \leq z\leq \gamma +\Delta \right. \right\} \\
 & \cup  & \left\{ (x,y,z)\in G^{+--}\left| \gamma \leq z\leq \gamma +\Delta \right. \right\} \\
 & \cup  & \left\{ (x,y,z)\in G^{---}\left| \gamma \leq z\leq \gamma +\Delta \right. \right\} \\
A^{+}_{\alpha }\wedge B^{+}_{\beta }\wedge C^{+}_{\gamma } & = & \left\{ (x,y,z)\in G^{+++}\left| \begin{array}{c}
\alpha \leq x\leq \alpha +\Delta \\
\beta \leq y\leq \beta +\Delta \\
\gamma \leq z\leq \gamma +\Delta 
\end{array}\right. \right\} \\
A^{-}_{\alpha }\wedge B^{+}_{\beta }\wedge C^{+}_{\gamma } & = & \left\{ (x,y,z)\in G^{-++}\left| \begin{array}{c}
\alpha \leq x\leq \alpha +\Delta \\
\beta \leq y\leq \beta +\Delta \\
\gamma \leq z\leq \gamma +\Delta 
\end{array}\right. \right\} \\
\vdots  &  & \\
A_{\alpha } & = & \left\{ (x,y,z)\in G^{+++}\left| \alpha \leq x\leq \alpha +\Delta \right. \right\} \\
 & \cup  & \left\{ (x,y,z)\in G^{-++}\left| \alpha \leq x\leq \alpha +\Delta \right. \right\} \\
 &  & \vdots \\
 & \cup  & \left\{ (x,y,z)\in G^{---}\left| \alpha \leq x\leq \alpha +\Delta \right. \right\} \\
\vdots  &  & \\
C_{\gamma } & = & \left\{ (x,y,z)\in G^{+++}\left| \gamma \leq z\leq \gamma +\Delta \right. \right\} \\
 & \cup  & \left\{ (x,y,z)\in G^{-++}\left| \gamma \leq z\leq \gamma +\Delta \right. \right\} \\
 &  & \vdots \\
 & \cup  & \left\{ (x,y,z)\in G^{---}\left| \gamma \leq z\leq \gamma +\Delta \right. \right\} 
\end{eqnarray*}
One can easily verify that the required symmetries (\ref{eq_S1}) and (\ref{eq_S2})
are satisfied by the following \emph{Ansatz}:\begin{eqnarray}
\rho ^{+++}(x,y,z)=\rho ^{--+}(x,y,z)=\rho ^{-+-}(x,y,z)  \nonumber  \\
 =\rho ^{+--}(x,y,z)= f(x+y+z)\rho (x+y+z)\label{eq_construct1} \\
\rho ^{-++}(x,y,z)=\rho ^{+-+}(x,y,z) =\rho ^{++-}(x,y,z)  \nonumber \\
 =\rho ^{---}(x,y,z)= \left( \frac{1}{4}-f(x+y+z)\right) \rho (x+y+z)\label{eq_construct2} 
\end{eqnarray}
where \( \rho  \) and \( f \) are arbitrary non-negative functions satisfying
the following conditions:\begin{eqnarray}
 & \int _{0}^{2\pi +\Delta }\int _{0}^{2\pi +\Delta }\int _{0}^{2\pi +\Delta }\rho (x+y+z)dxdydz=1 & \label{eq_construct5} \\
 & 0\leq f(w)\leq \frac{1}{4}\, \, \, \, \, \, \, w\in [0,6\pi  ] & \label{eq_construct6} 
\end{eqnarray}

Now, the probability measure on the hidden variable space, that is, the functions
\( \rho  \) and \( f \) must be defined in such a way that the quantum probabilities
(\ref{eq_prob1})-(\ref{eq_prob3}) are reproduced. Due to (\ref{eq_construct1})-(\ref{eq_construct6}),
equation (\ref{eq_prob1}) is automatically satisfied, and if \( \rho  \) and
\( f \) satisfy (\ref{eq_prob2}) then they automatically satisfy (\ref{eq_prob3}). 
So, there remains only one equation to be solved: \begin{eqnarray}
\label{eq_tobesolved}
\frac{\int _{\gamma }^{\gamma +\Delta }\int _{\beta }^{\beta +\Delta }\int _{\alpha }^{\alpha +\Delta }f(x+y+z)\rho (x+y+z)dxdydz}{\int _{\gamma }^{\gamma +\Delta }\int _{\beta }^{\beta +\Delta }\int _{\alpha }^{\alpha +\Delta }\rho (x+y+z)dxdydz}\nonumber\\=\frac{1}{8}\left( 1-\cos \left( \alpha +\beta +\gamma \right) \right) 
\end{eqnarray}
So we are looking for non-negative real functions \( \rho (w) \) and \( f(w) \)
defined on the interval \( \left[ 0,6\pi \right]  \), satisfying (\ref{eq_tobesolved})
and the conditions (\ref{eq_construct5}) and (\ref{eq_construct6}). We know that \( f(w)=0 \) if \( w\in \bigcup _{k=0,1,2,3}\left[ 2k\pi ,2k\pi +3\Delta \right]  \)
because \( \cos \left( 2k\pi \right) =1,\, \, k=0,1,2,3 \), and \( f(z)=\frac{1}{4} \)
if \( w\in \bigcup _{k=0,1,2}\left[ (2k+1)\pi ,(2k+1)\pi +3\Delta \right]  \)
because \( \cos \left( (2k+1)\pi \right) =-1,\, \, k=0,1,2 \).  These two regions
must be disjoint, consequently \( \Delta \leq \frac{\pi }{3} \), which means
-- in this model -- a limitation for the single detection efficiency: \( \omega \leq \frac{1}{6} \).
Let us chose \( \Delta =0.9\frac{\pi }{3} \), $\omega=15\%$.
\begin{figure}[t]
	\begin{center}\leavevmode
	\epsfxsize=5.5cm
	\epsfbox{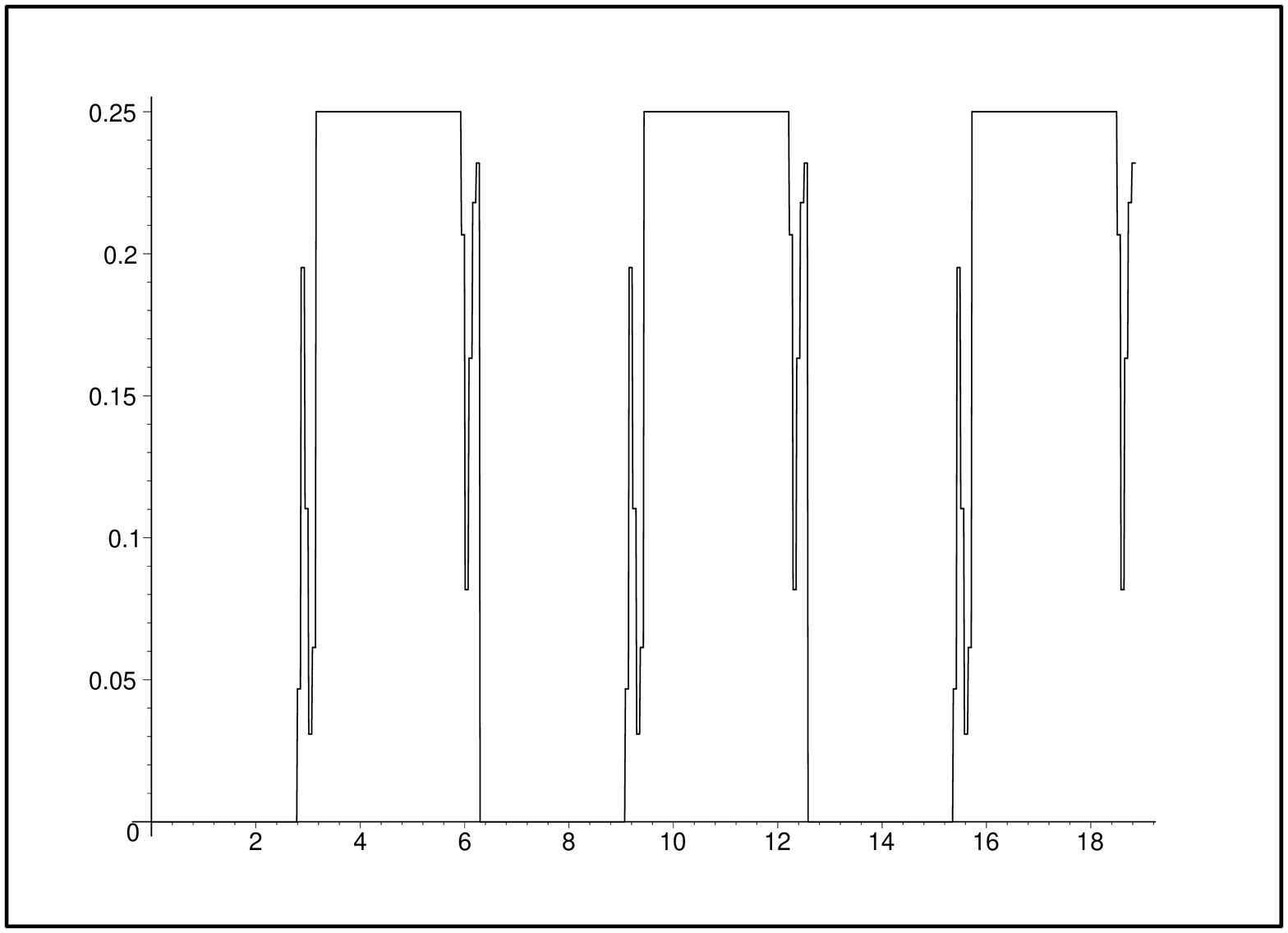}
	\end{center}
	\caption{A numerical solution of the integral equation (\ref{eq_tobesolved}) for function \protect\( f\protect \)}
\label{fig_f} 
\end{figure} 
\begin{figure}[t]
  	\begin{center}\leavevmode
	\epsfxsize=5.5cm
	\epsfbox{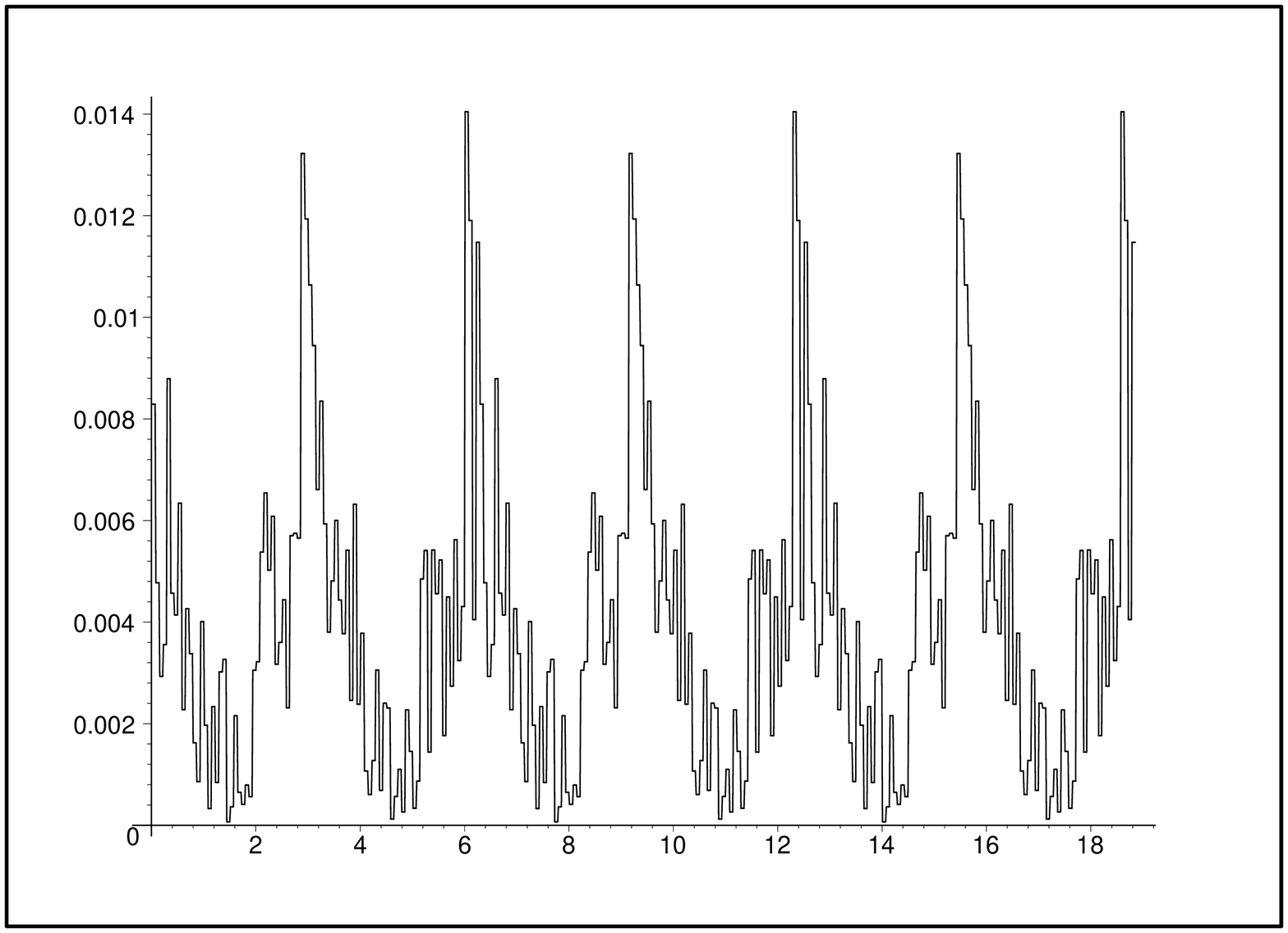}
  	\end{center}
	\caption{A numerical solution of the integral equation (\ref{eq_tobesolved}) for function \protect\( \rho \protect \)}
\label{fig_measure}
\end{figure}
\begin{figure}[t]
	\begin{center}\leavevmode
	\epsfxsize=5.5cm
	\epsfbox{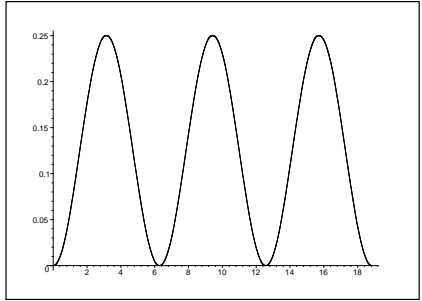}
	\end{center}
	\caption{This is two curves in coincidence: One is the numeric result  of the left hand side of equation (\ref{eq_tobesolved}), the other is the function on the right hand side}
\label{fig_fit}
\end{figure}
\begin{figure}[t]
	\begin{center}\leavevmode
	\epsfxsize=5.5cm
	\epsfbox{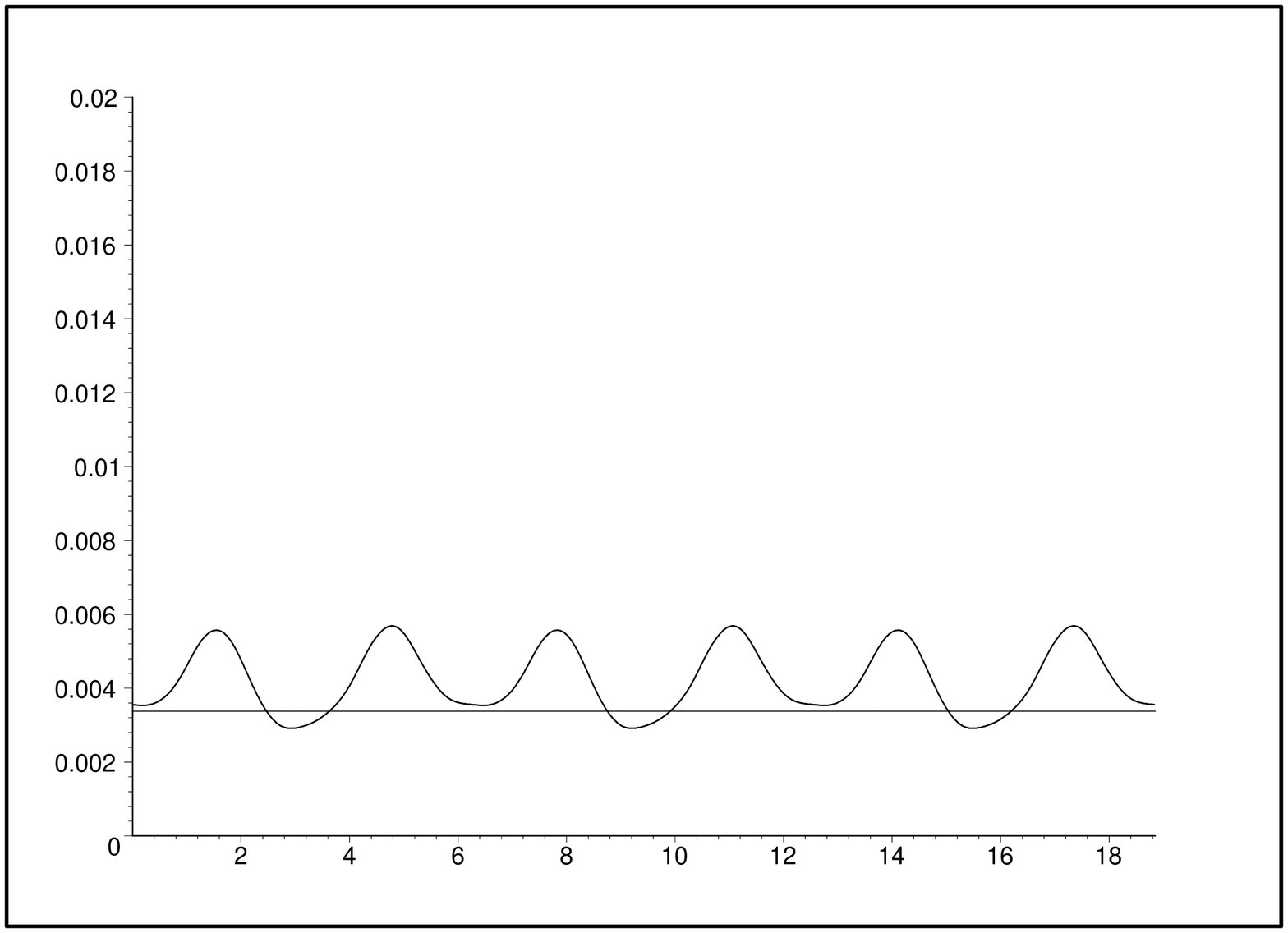}
	\end{center}
	\caption{The dependence of the triple detection efficiency on \protect\( \alpha +\beta +\gamma \protect \). The horizontal line represents the value \protect\( \omega ^{3}=0.34\%\protect \)
corresponding to the case of independence}
\label{fig_triple}
\end{figure}

One can solve (\ref{eq_tobesolved}) numerically. Figure~\ref{fig_f}
 and \ref{fig_measure} show the numerical solutions for functions $f$ and \( \rho  \). Figure~\ref{fig_fit}
illustrates the high precision of the numerical solution.  
The triple detection efficiency depends on the phase shift angles.
Figure~\ref{fig_triple} shows the dependence of the triple detection efficiency
on \( \alpha +\beta +\gamma  \). The minimal triple efficiency -- in this example
-- is about 0.2\%.

\section{Recent experiments}
\begin{figure}[h]
	\begin{center}\leavevmode
	\epsfxsize=6cm
	\epsfbox{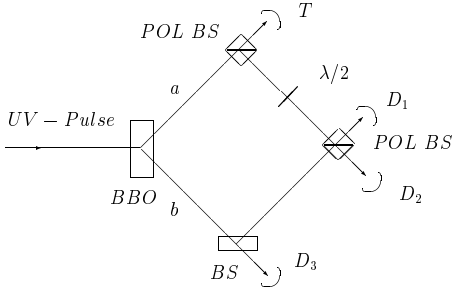}
	\end{center}
	\caption{ The experimental setup for demonstration of GHZ entanglement for
spatially separated photons}\label{fig_recent}
\end{figure}

Figure~\ref{fig_recent} shows the schematic drawing 
of the experimental setup
of the Innsbruck experiment \cite{Z}. With a small probability,
an
UV pulse causes a double pair creation in the non-linear crystal (BBO).
The
two pairs created within the window of observation are indistinguishable.
It
can be shown that by restricting the ensemble to the sub-ensemble of cases
when 
all of the four detectors, \( T,D_{1},D_{2},D_{3} \) fire, we obtain the
following
quantum state:
\[
\underbrace{\frac{1}{\sqrt{2}}\left( \left| H\right\rangle _{1}\otimes
\left| H\right\rangle _{2}\otimes \left| V\right\rangle _{3}+\left|
V\right\rangle _{1}\otimes \left| V\right\rangle _{2}\otimes \left|
H\right\rangle _{3}\right) }_{\Psi _{GHZ}}\otimes \left| H\right\rangle
_{T}\]
where \( \left| H\right\rangle _{T} \) denotes the state of the
photon at detector \( T \). This quantum state corresponds to a
four-particle system consisting of an entangled three-photon system
in GHZ state, and a fourth independent photon. So we may assume that
the statistics observed on the sub-ensemble conditioned by the
four-fold coincidences are the same as those taken on the
sub-ensemble conditioned by the triple detections at \( D_{1},D_{2}
\) and \( D_{3} \).  What is important from our point of view is that
\emph{any further experimental observations testing the GHZ
correlations, which are based on the above described preparation of
GHZ entangled states, will be performed on selected sub-ensembles
conditioned by the triple coincidence detections. Therefore, all of
these experimental observations will be treated by our local hidden
variable model}.

Finally, notice that triple detections, where permitted by the prism model,
are subject to ordinary sorts of external detection error. If the external
detection efficiency is, say, $d$, then triple outcomes having probability
\[ p(\mathrm{triple\ detection}|\mathrm{none\ are\ defective})=1 \] according
to the ideal case specified in the model, will have a reduced probability
of $d^3$, as in the  usual analysis of random errors. Similarly we can take
into account the non-zero probability of random dark photon detections and
make a calculation like that of de Barros and Suppes, resulting in the
modified expectation values (\ref{expectation3}) and (\ref{expectation4}). 
Thus our local hidden variable framework allows for the
usual techniques of error analysis to treat experimental inefficiencies
reflected in the actual observations. 

\vskip0.5cm
The research was partly supported by the OTKA Foundation, No.~T025841 and
No.~T032771 (L. E. Szab\'o).


\end{document}